\documentstyle[sprocl]{article}

\def\curl{\mathop{\rm curl}\nolimits}
\newcommand{\df}{{\mbox{\rm d}}}
\newcommand\bm[1]{\mbox{\boldmath $#1$}}
\newcommand{\smfrac}[2]{{\textstyle{#1\over#2}}}
\def\half{\smfrac{1}{2}}

\begin{document}

\title{INTEGRABILITY IN TETRAD FORMALISMS AND CONSERVATION IN COSMOLOGY}

\author{M.A.H. MacCALLUM}

\address{School of Mathematical Sciences \\
Queen Mary and Westfield College, University of London \\
Mile End Road, London E1 4NS, U.K. \\
E--mail: M.A.H.MacCallum@qmw.ac.uk}

\maketitle\abstracts{Integrability in general tetrad formalisms is
reviewed, following and clarifying work of Papapetrou and Edgar. The
integrability conditions are (combinations of) the Bianchi equations
and their consequences. The introduction of additional constraints is
considered. Recent results on the conservation of constraints in the
1+3 covariant formulation of cosmology are shown to follow from the Bianchi
equations.}

\section{Introduction}

In a number of recent works,\cite{Els96,Maa97,MaaLesEll97,Vel97}
calculations have been made to show that when the governing equations
of relativistic cosmology, in the covariant 1+3 formalism, are divided
into evolution equations and constraints, the constraints are
preserved under the evolution. Unfortunately these attempts do not
wholly agree with one another, though all agree the constraints are
conserved.

On general principle, the only integrability equations that should
arise are the Bianchi identities (regarded as equations rather than
identities, in tetrad formalisms) and their consequences, and the
derivatives of any additional constraints imposed (e.g.\ the perfect
fluid form). Thus the cited results on the evolution of constraints
should follow immediately from the Bianchi equations. Here we show
explicitly that this is true, noting some subtleties involved.

Integrability for general tetrad formalisms has been discussed by
Edgar\cite{Edg80}, following work of
Papapetrou\cite{Pap70,Pap71,Pap71a}. The first section of this paper
reviews and in part clarifies that treatment.  Section 2 considers the
introduction of additional constraints and section 3 shows how the
conservation of the first two of the cosmological constraints
previously discussed can be derived from the Bianchi equations, and
reconciles this with previous treatments. A fuller discussion of these
issues, in particular completing the discussion of the 1+3
constraints, will appear elsewhere.

\section{Integrability in tetrad formalisms}

In the Cartan approach to curvature, there are two basic
structure equations for a set of basis one-forms $\bm{\omega}^a$. For
a Riemannian space these are\footnote{Index and sign
conventions here follow Kramer et al.\cite{KraSteMac80} unless otherwise
stated.}
\begin{equation}
\df \bm{\omega}^a=-\bm{\Gamma}^a{}_b\wedge \bm{\omega}^b \label{cartan1}
\end{equation}
which utilises the connection one-forms $\bm{\Gamma}^a{}_b$, and
\begin{equation}
\df \bm{\Gamma}^a{}_b+\bm{\Gamma}^a{}_c\wedge \bm{\Gamma}%
^c{}_b=\bm{R}^a{}_b \label{cartan2}
\end{equation}
which defines the curvature 2-forms $\bm{R}^a{}_b$. The integrability
conditions of these equations are simply the conditions $\df^2
\bm{\omega}^a = 0$ and $\df^2 \bm{\Gamma}^a{}_b =0$, the first and
second Bianchi identities, the first of which can also be considered
to be the Jacobi identities for a triple of the basis vectors dual to
the $\bm{\omega}^a$,
\begin{equation}
J^a{}_{bcd} \equiv
\bm{\omega}^a\left(\left[\bm{e}_b,\left[\bm{e}_c,\bm{e}_d\right]\right] +
\left[\bm{e}_c,\left[\bm{e}_d,\bm{e}_b\right] \right]+
\left[\bm{e}_d,\left[\bm{e}_b,\bm{e}_c\right]\right]\right)=0~. \label{Jacobi}
\end{equation}

Written in terms of forms, these Bianchi identities read
\begin{eqnarray}
\bm{R}^a{}_c \wedge \bm{\omega}^c &=&0 ~,\label{Bianchi1}\\
\df \bm{R}^a{}_b - \bm{R}^a{}_c \wedge \bm{\Gamma}^c{}_b + \bm{\Gamma}^a{}_c
 \wedge \bm{R}^c{}_b&=&0~. \label{Bianchi2}
\end{eqnarray}
Note that (\ref{cartan2}) and (\ref{Bianchi2}) guarantee the
integrability of (\ref{Bianchi2}).

Tetrad formulations normally start (though
they need not) with a basis in which the metric is fixed, i.e.\ if
$g_{ab}$ is the scalar product of $\bm{e}_a$ and $\bm{e}_b$,
\begin{equation}
\df g_{ab} = \df (\bm{e}_a . \bm{e}_b) = 0 ~. \label{metricity}
\end{equation}
Tetrad indices can be lowered and raised with $g_{ab}$ and its inverse.
Together with (\ref{cartan1}), which can also be written as the
commutator equations
\begin{equation}
[\bm{e}_a, \bm{e}_b]=-2{\Gamma}^{c}{}_{[ab]}\bm{e}_{c}, \label{commut}
\end{equation}
and fixes the skew parts $\Gamma^a{}_{[bc]}$ of the coefficients in
$\bm{\Gamma}^a{}_b=\Gamma^a{}_{bc}\bm{\omega}^c$, (\ref{metricity}) determines
$\Gamma^a{}_{bc}$. Thus, under the condition (\ref{metricity}),
(\ref{commut}) is equivalent to the definition 
\begin{equation}
\Gamma ^c{}_{ab}\equiv \omega ^c{}_ke_b{}^ie_a{}^k{}_{;i}=-\omega
^c{}_{k;i}e_b{}^ie_a{}^k \label{gammadef}
\end{equation}
in terms of the cooordinate components (indices $i,\,j,\,k,\dots$) of
the basis vectors and one-forms. Papapetrou and Edgar both remark on
this equivalence, but without explicitly emphasizing that it only
holds when (\ref{metricity}) is assumed.

For a Riemannian space, the
components $R^a{}_{bcd}$ of the curvature two-form
\begin{equation}
\bm{R}^a{}_b=\half R^a{}_{bcd}\bm{\omega}^c \wedge  \bm{\omega}^d
\end{equation}
must satisfy
\begin{equation} 
R_{abcd}=-R_{bacd}=-R_{abdc}=R_{cdab},~~ R_{a[bcd]} = 0~.\label{riesymm}
\end{equation}

In a tetrad formulation, it is usual to regard $g_{ab}$ as given and
the components $e_b{}^i$, $\Gamma^c{}_{ab}$ and $R^a{}_{bcd}$ as
functions to be solved for. The equations for them are then the tetrad
component forms of (\ref{cartan1}), (\ref{cartan2}) and
(\ref{Bianchi2}). Since (\ref{Bianchi2}) is now regarded as an
equation, one has to adjoin its integrability condition to close the
system. Note that it is assumed that the components $\Gamma^c{}_{ab}$
satisfy the symmetry condition $\Gamma_{cab}+\Gamma_{acb}=0$ and that
the $R^a{}_{bcd}$ satisfy (\ref{riesymm}):  both these are conditions
which are implicit in the usual discussions, but they are important
and should be made explicit.

The equations of the tetrad formalism can thus (always assuming the given
constant metric and the above symmetry conditions) be written as
\begin{eqnarray}
\hspace{-1.4em}
0 &= \delta \Gamma^a{}_{bc\phantom{de}} &= \Gamma ^c{}_{[ab]}- \omega
 ^c{}_ke_{[b}{}^ie_{a]}{}^k{}_{,i}~, \label{delgam}\\
\hspace{-1.4em}
0 &= \delta R_{abcd\phantom{e}} &=
 R_{abcd}-\Gamma_{abd,c}+\Gamma_{abc,d} - \Gamma^e{}_{bd}\Gamma_{aec} +
 \Gamma^e{}_{bc}\Gamma_{aed} -2\Gamma^e{}_{[cd]}\Gamma_{abe},
  \label{delrie} \\
\hspace{-1.4em}
0 &= \delta B_{abcde} &= R_{abcd;e} + R_{abde;c} + R_{abec;d}~. \label{delbian}
\end{eqnarray}
Here a comma denotes a partial or directional derivative, and
semi-colon denotes the covariant derivative for which the
$\Gamma^a{}_{bc}$ are the connection coefficients (and which is the
metric connection if (\ref{delgam}) and (\ref{metricity}) hold); this
differs from Edgar's use of the semi-colon for a directional
derivative of a tetrad component. The Bianchi identity (\ref{delbian})
has of course to be expanded in terms of directional derivatives and
the $\Gamma^a{}_{bc}$ in order to give a set of equations entirely in terms of
the declared set of variables.

The quantity $\delta R_{abcd}$ in (\ref{delrie}) does not obey
(\ref{riesymm}) though it does satisfy
\begin{equation} 
\delta R_{abcd}=-\delta R_{bacd}=-\delta R_{abdc}~.\label{riesymm1}
\end{equation}
In 4 dimensions, there are six distinct skew pairs of indices so a
tensor obeying (\ref{riesymm1}) has 36 distinct components. For
$\delta R_{abcd}$ these 36 
can be decomposed into a
Riemann tensor part with the symmetries (\ref{riesymm}) and 20
independent components which can be further split into the Weyl and
Ricci tensor parts $\delta C_{abcd}$ and $\delta R_{ab}$, a part
$\delta R_{abcd}-\delta R_{cdab}$ (15 components) and a part $\delta
R_{[abcd]}$ (1 component). The last two parts give $\delta J^a{}_{bcd}$.

These equations are not exactly Cartan's, due to
substitutions and the taking of tetrad components, so the integrability
conditions are combinations of the Bianchi equations.  From Edgar's
calculations, transcribed to my notation, these integrability conditions are
\begin{eqnarray}
\hspace{-1em}
0 &=& \delta J^a{}_{bcd} - \delta \Gamma^a{}_{[bd,c]}
- \delta \Gamma^a{}_{[b|e|} \delta \Gamma^e{}_{cd]}
- \delta \Gamma_{e[bc} \delta \Gamma^{ea}{}_{d]}
+ \delta \Gamma_{e[bc} \Gamma^{ea}{}_{d]} \nonumber \\
\hspace{-1em}
&& 
+ \delta \Gamma^a{}_{[b|e|} \Gamma^e{}_{cd]}
+ \delta \Gamma^a{}_{e[c} \Gamma^e{}_{bd]}
+ \delta \Gamma^e{}_{[cd} \Gamma^a{}_{b]e}~, \label{int1}\\
\hspace{-1em}
0 &=& \delta B_{abcde} - (\delta R_{abcd}){}_{;e} - (\delta
      R_{abde}){}_{;c} - (\delta R_{abec})_{;d} \nonumber \\ 
\hspace{-1em}
&&
     - 6 \Gamma_{ab[c|,g|}\delta\Gamma^g{}_{de]} -3\Gamma_{ab}{}^{g}
      \delta J_{gcde}~, \label{int2} \\
\hspace{-1em}
0 &=&\eta^{cdef}((\delta B_{abcde}){}_{;f} - 3R_{abgc}\delta
J^g{}_{def}
- 3\delta R^g{}_{[a|cd}R_{g|b]ef}
-3R_{abcd,g}\delta\Gamma^g{}_{ef}. \label{int3}
\end{eqnarray}
One can check that no further conditions arise.

These results show that there is some redundancy in
(\ref{delgam})-(\ref{delbian}), e.g.\ that if
(\ref{delgam})-(\ref{delrie}) are satisfied so are
(\ref{int1})-(\ref{int3}) and (\ref{delbian}). However, it is
important, as Edgar has noted, that (\ref{delrie})-(\ref{delbian}) are
not a complete set of equations. This, and misunderstandings about the
role of (\ref{metricity}) and (\ref{commut}), have contributed to
uncertainty about whether the tetrad equations form a complete system
and if so which equations are required (see for example\footnote{This
reference was drawn to my attention by Prof.\ B.\ Edgar.}
Chandrasekhar\cite{Cha83}). The situation has been discussed by
Edgar\cite{Edg80,Edg92}. 

One could take a different set of variables. For instance, one could
adjoin functions supposed to be the values of $R_{abcd;e}$; then
(\ref{Bianchi2}), or more precisely the definition of $R_{abcd;e}$ in terms
of $R_{abcd}$ and $\Gamma^a{}_{bc}$, are algebraic equations for
$R_{abcd;e}$, and (\ref{int3}) is a differential equation, giving
further integrability conditions. Finally (as Dr.\ A. Rendall has
reminded me) one should note that in general integrability conditions
are necessary but not sufficient to guarantee existence and uniqueness
of solutions, which depend also on analytic properties of the
equations and data.

\section{Additional constraints}

The equations above assume a completely general spacetime and choice
of tetrad. Additional equations and integrability conditions arise if
one imposes additional conditions which amount to replacing some of
the inequalities which must hold for the general case by
equalities. Among these, for equations of state which limit the
energy-momentum tensor, are Einstein's field equations. Edgar reminds
us that the above tetrad equations remain consistent in the case of
vacuum in general relativity. For the cosmological case this is also
true for (for instance) perfect fluids.

One could distinguish between two different cases of choice of
equality, or set of equalities, as follows.

First, one could choose a tetrad in a way that is compatible with the
general case under study. 
The derivatives of this constraint will then give
additional conditions. However, the processes of imposing 
the constraint
and solving the equations must commute; that is to say one could in
principle have solved the equations in a general basis and then chosen
the tetrad 
by applying a Lorentz transformation at
each point (i.e.\ algebraically). Therefore the extra equations in
this case must be just the differential equations necessarily
satisfied by a tetrad chosen {\em after} solving the original tetrad
equations.
%
%
In the cosmological context an example of a condition of this type is
given (assuming the energy-momentum obeys the usual energy conditions
pwhich guarantee the existence of a timelike eigenvector of the Ricci
tensor) by the requirement of zero heat flux, $q_a=0$ in the usual
notation\cite{Ehl61,Ell71}.

A second type of equality or set of equalities that can be imposed is
one which gives a consistent specialization of the general system of
equations but which is not generally possible just by choice of
tetrad: for example, the conditions that a spacetime is vacuum, of
Petrov type D, and has shearfree and geodesic principal null
congruences.
These can, at least intuitively, be
regarded as specifications of subspaces, in some suitable function
space, such that solutions of the tetrad equations which start in the
subspaces lie wholly in them. In cosmology, the assumption that the
matter content is irrotational `dust' (with no acceleration) is such a
set of conditions.

 One often wants to work out such a set starting from some incomplete
set of conditions. An example is given by the investigations, in
cosmology, aimed at deciding what solutions, if any, satisfy
conditions such as the vanishing of the electric or magnetic parts of
the Weyl tensor\cite{MaaLesEll97,ElsUggLes97,Sop97}. This can become very
confusing when the conditions are substituted into the general tetrad
equations and integrability conditions are then considered. It seems
to me, though I have not so far found a proof, that one can always
arrive at the consequences in a clearer way by taking the extra
conditions, say $C^A=0$ for some set of indices $A$, and just adding
the equations $C^A{}_{,ab\ldots}=0$ until the system closes, rather
than intermixing the new conditions with the original equations before
constructing the integrability conditions for them.

\section{Application to the 1+3 covariant formalism in cosmology}

The approach referred to here is the one\cite{Ehl61,Ell71} based on a
timelike congruence with unit tangent vector $u^a$: this is a
`threading' rather than `slicing' approach\cite{JanCarBin92}. It is
covariant in the sense that it is coordinate-independent, and
invariant if $u^a$ is invariantly-defined. The metric is taken to have
signature +2 and one defines $h_{ab}=g_{ab}+u_a u_b$.  The kinematic
quantities $\Theta$ (expansion), $\sigma_{ab}$ (shear), $\omega_{ab}$
(vorticity) and $\dot{u}^a$ (acceleration) are then defined by
\begin{equation}
u_{a;b}= \smfrac{1}{3}\Theta h_{ab} + \sigma_{ab} +
\omega_{ab}-\dot{u}_a u_b
\end{equation}
where
\begin{equation}
\sigma_{ab}=\sigma_{ba}, ~\sigma^a{}_a = 0, ~\sigma_{ab} u^b =
0,~\omega_{ab}=-\omega_{ba}, ~\omega_{ab} u^b = 0~.
\end{equation}

From the point of view above, one can regard this as an incomplete orthonormal
tetrad formalism in which the spatial triad has not been specified
(but in principle could be, if $u^a$ is invariantly defined, by
following the methodology of the approach to characterization of
spacetimes due to Cartan and Karlhede: see e.g.\ MacCallum and Skea%
\cite{MacSke94}). In this way of looking at the method, the kinematic
quantities are (some of) the connection coefficients
$\Gamma^a{}_{bc}$.

The energy-momentum tensor is split by
\begin{equation}
G_{ab} \equiv R_{ab}-\half R g_{ab} = T_{ab} = \mu u_a u_b +2q_{(a}
u_{b)} + ph_{ab} + \pi_{ab}
\end{equation}
where $q_a u^a = 0$ ,$\pi_{ab}=\pi_{ba}$, $\pi^a{}_a =0$ and $\pi_{ab}
u^b = 0$. The Weyl tensor
$C_{abcd}$ can be written in terms of electric and magnetic parts,
$E_{ab}$ and $H_{ab}$, which are both symmetric traceless tensors
orthogonal to $u^a$, as
\begin{eqnarray}
C_{abcd} &=&\bar{g}_{ac}E_{bd}-\bar{g}_{ad}E_{bc}
-\bar{g}_{bc}E_{ad}+\bar{g}_{bd}E_{ac}\nonumber\\ 
&&+\eta_{abe}u_cH_d{}^e - \eta_{abe}u_dH_c{}^e + \eta_{cde}u_aH_b{}^e
-\eta_{cde}u_bH_a{}^e  \label{weyldecomp}
\end{eqnarray}
where $\bar{g}_{ab} = g_{ab}+2u_a u_b$ and $\eta_{abc} = \eta_{abcd}
u^d$ is a three-dimensional volume element (in the tangent space at a
point in the general case where $\omega_{ab} \neq 0$ and hence the
planes defined by $h_{ab}$ are not surface-forming). Later I will
refer to (combinations of) the equations (\ref{delrie}), etc., as
$\delta q_a$ and so on.
 
One now proceeds to split the equations (\ref{delrie}-\ref{delbian}),
as far as possible, by invariant projections parallel and perpendicular
to $u^a$, ignoring those equations which cannot be written in terms of
the above variables. The equations are separated into evolution
equations, giving values of derivatives $\dot{Q} = Q_{;a}u^a$ of
quantities $Q$, and constraints which do not involve such
derivatives. Note that for a tensor projected orthogonal to $u^a$ the
projection of $\dot{Q}$ differs from $\dot{Q}$ only by multiples of
$u^a$ and $\dot{u}^b$ with $Q$, so when studying conservation it is
sufficient to study the projection.

In the general case, the evolution equations consist of three Jacobi
identities giving the evolution of $\omega_{ab}=\eta_{abc}\omega^c$, a
combination of the Einstein equations, the Raychaudhuri equation, giving
$\dot{\Theta}$, five components of $\delta R_{abcd}$ giving
$\dot{\sigma}_{ab}$, and 14 Bianchi equations giving $\dot{\mu}$,
$\dot{q}_a$, $\dot{E}_{ab}$ and $\dot{H}_{ab}$. The constraints
consist of one Jacobi identity giving $\omega^a{}_{;a}$, three
Einstein equations giving $q_a$, five components of $\delta R_{abcd}$
giving $H_{ab}$ and 6 components of $\delta B_{abcde}$ giving $D^aE_{ab}$
and $D^aH_{ab}$, where the operator $D^a$ is defined for any tensor by
\begin{equation}
D_{a}T^{c\ldots d}{}_{e\ldots f} = h_a{}^bh^c{}_p\ldots h^d{}_q
h_e{}^r\ldots h_f{}^s T^{p\ldots q}{}_{r\ldots s;b}~.
\end{equation}
The general version of the tetrad equations is not needed in full in
this paper, and so is omitted. Note that the evolution of some
quantities remains undetermined and depends on, for example, an
equation of state of matter.

We note that in this system, compared with a general tetrad system, we
have omitted 12 Jacobi equations and 6 components of
$\delta R_{abcd}$. The latter are those field equations, in a tetrad
whose spacelike basis vectors are indicated by use of Greek sub- and
superscripts, which require $R_{\alpha\beta}$. Without these, of
course, the system is incomplete, and in particular the
$\dot{\sigma}_{ab}$ equations require a combination of $E_{ab}$ and
$\pi_{ab}$ rather than these two, as might seem more natural, being
separated. When one can make a separation, i.e.\ when one has a complete
tetrad, it is usual to distinguish the field equations
$G_{\alpha\beta}=T_{\alpha\beta}$ which give
$\dot{\sigma}_{\alpha\beta}$ and equations for $E_{ab}$ and
$G_{ab}u^a u^b$ which are constraints. This organization of the
possible linear combinations of the equations is especially
appropriate when considering a Cauchy problem, since the constraints
$G_{ab}u^b = T_{ab}u^b$ are conserved by the evolution equations
$G_{\alpha\beta}=T_{\alpha\beta}$ as a consequence of the contracted
Bianchi identities, as is well-known.

The incompleteness of the system of equations without
$G_{\alpha\beta}=T_{\alpha\beta}$ or an equivalent is discussed by
Ellis\cite{Ell71} and Velden\cite{Vel97}. In the latter work, however,
the missing equations
themselves are not written down (they are only implicit in the
statement in words of the meaning of $R_{\alpha\beta}$) and the linear
combinations taken give $\dot{\sigma}_{ab}$ in terms of $E_{ab}$
(only, among the components of $R_{abcd}$) and constraints for
$R_{\alpha\beta}$. This system is then treated as a complete system of
evolution and constraint equations, although it is still missing the
equations giving $R_{\alpha\beta}$ in terms of the connection or
metric and is perhaps therefore better regarded, in the context of the
study of evolution (though not in some of Velden's other discussions)
as just adjoining to the usual system, with the equation $\delta
R_{\alpha\beta}=0$ omitted, some equations giving algebraic
definitions of additional quantities which would, if $R_{\alpha\beta}$
were in fact calculated, be that tensor. In a complete tetrad
formalism, the 12 extra Jacobi identities include the evolution
equations for the remaining $\Gamma^a{}_{bc}$ insofar as these are
fixed without conditions on the tetrad.

It is now my aim to show that the discussion of evolution of
constraints in the various works cited can be more simply derived, and
more deeply understood, by considering specialization of the
conditions (\ref{int1}-\ref{int3}). Here I will illustrate this
only for the irrotational dust case, and only for two of the
constraints discussed in previous works; a fuller account will appear
elsewhere.
For definiteness, I shall follow Maartens\cite{Maa97}, though the other
related papers use similar methods and notation. From the present
point of view, the method is to assume 
that $\delta \Gamma^a{}_{bc}=0=\delta J_{abcd}=\delta \pi_{ab}=
\delta p$ and that $\delta R_{abcd} = 0 =\delta B_{abcde}$ for the
components giving evolution equations. In the equations one can set
$\dot{u}^a =0 = \omega^c = q^a=\pi_{ab}=p$, which in particular means
all Jacobi identities are ignored. For reasons which will become
apparent, however, I will retain a non-zero $q^b$ for the moment. The
evolution equations are then
\begin{eqnarray}
 0 &=& \delta R_{00}=\dot{\Theta} + \smfrac{1}{3} \Theta^2 + \half \mu +
 \sigma_{ab}\sigma^{ab} ~,\nonumber \\
 0 &=& \delta(E_{ab}-\half\pi_{ab})= h_a{}^ch_b{}^d\dot{\sigma}_{cd}
 +\smfrac{2}{3}\Theta
 \sigma_{ab} + \sigma_{c<a}\sigma_{b>}{}^c + E_{ab} ~,\nonumber \\
 0 &=& \delta {\rm Bian}(\dot{\mu})=\dot{\mu} + \Theta \mu ~,\nonumber \\
 0 &=&\delta {\rm Bian}(\dot{q}_a) = h_a{}^b\dot{q}^b + \sigma_{ab}q^b
 +\smfrac{4}{3}\Theta q_a ~, \nonumber \\
 0 &=& \delta {\rm Bian}(\dot{E}_{ab})= h_a{}^ch_b{}^d\dot{E}_{cd} +
 \Theta E_{ab} -
 3\sigma_{c<a}E_{b>}{}^c -
 \eta_{(a|cd|}D^cH_{b)}{}^d -\half \mu \sigma_{ab} ~,\nonumber \\
 0 &=& \delta {\rm Bian}(\dot{H}_{ab})=h_a{}^ch_b{}^d\dot{H}_{cd} +
 \Theta H_{ab} - 
 3\sigma_{c<a}H_{b>}{}^c + \eta_{(a|cd|}D^cE_{b)}{}^d ~, \nonumber
\end{eqnarray}
where I have adopted the convention that angle brackets denote the
traceless symmetric part. The names $\delta R_{00}$ and so on have
been introduced to make the following discussion easier to read
without referring back to numbered equations.

The constraints are
\begin{eqnarray}
 0 &=&\delta q_a \equiv q_a+D^b \sigma_{ab} -\smfrac{2}{3}D_a \Theta
 ~,\nonumber \\
 0 &=&\delta H_{ab} \equiv H_{ab} - \eta_{(a|cd|}D^c\sigma_{b)}{}^d
 ~,\nonumber \\
 0 &=& \delta {\rm Bian}(D^bE_{ab}) \equiv D^bE_{ab} - \smfrac{1}{3}D_a \mu
 - \eta_{abc} \sigma^b{}_d H^{cd} -\half \sigma_{ab}q^b
 +\smfrac{1}{3}\Theta q_a~,\nonumber \\
 0 &=& \delta {\rm Bian}(D^bH_{ab}) \equiv D^bH_{ab}+ \eta_{abc}
 \sigma^b{}_d E^{cd} - \half \eta_{cd(a}\sigma^c{}_{b)}q^d~.\nonumber
\end{eqnarray}

From $\delta {\rm Bian}(\dot{q}_a)$ the evolution
equation for $\delta q_a$ should be
\begin{equation}
 0 = h_a{}^b\dot{\delta q}_b + \sigma_{ab}\delta q^b
 +\smfrac{4}{3}\Theta \delta q_a ~.
\end{equation}
However, to calculate, from the above system, the evolution of $\delta q_a$
(denoted ${\cal C}^1_a$ by Maartens\cite{Maa97}), one has to
eliminate $\dot{\Theta}$ and $\dot{\sigma}_{ab}$ using $\delta
R_{00}=0$ and $\delta(E_{ab}-\half\pi_{ab})=0$, which, from the Bianchi identities,  naturally brings
in $\delta {\rm Bian}(D^bE_{ab})$ (the
precise combination is given by a component of
(\ref{int2})). Thence we should have
\begin{equation}
 0 = h_a{}^b\dot{\delta q}_b + \sigma_{ab}\delta q^b
 +\smfrac{4}{3}\Theta \delta q_a  + \delta {\rm Bian}(D^bE_{ab})~,
\end{equation}

Since we are assuming the $\Gamma^a{}_{bc}$ are
correct, the initially zero $q_a$ has to be given the value $-\delta
q_a$ while calculating the evolution of $\delta q_a$, and similarly
the initial $H_{ab}$ has to be replaced by $H_{ab} - \delta
H_{ab}$ ($\delta H_{ab}$ is denoted $-{\cal C}^2_{ab}$ by Maartens%
\cite{Maa97}). Thus if ${\cal C}^3_a = D^bE_{ab} - \smfrac{1}{3}D_a
\mu - \eta_{abc} \sigma^b{}_d H^{cd}$, then $\delta {\rm Bian}(D^bE_{ab}) =
{\cal C}^3_a + \eta_{abc} \sigma^b{}_d \delta H^{cd}+\half
 \sigma_{ab}\delta q^b -\smfrac{1}{3}\Theta \delta q_a$ leading to
\begin{equation}
 0 = h_a{}^b\dot{\delta q}_b + \smfrac{3}{2} \sigma_{ab}\delta q^b
 +\Theta \delta q_a  + {\cal C}^3_a -\eta_{abc} \sigma^b{}_d {\cal C}^{2cd}~,
\end{equation}
This is precisely Velden's result. Maartens\cite{Maa97} gives
\begin{equation}
 0 = h_a{}^b\dot{\delta q}_b +\Theta \delta q_a  + {\cal C}^3_a
 -2\eta_{abc} \sigma^b{}_d {\cal C}^{2cd}~,
\end{equation}
The discrepancy arises because the modified Ricci identities with
non-zero $\delta R_{abcd}$ have only been used in some of the places
where they should appear. Since the $\Gamma^a{}_{bc}$ are taken to be
correct, the $\delta R_{abcd}$ terms have to be added to the
commutators wherever the corresponding $R_{abcd}$ appear. For
example, one should use
\begin{equation}
(D^bS_{ab})\dot{} = D^b\dot{S}_{ab} - \smfrac{1}{3}\Theta D^bS_{ab} -
\sigma^{bc}D_cS_{ab}+\eta_{abc}(H^b{}_d +{\cal
C}^{2b}{}_d)S^{cd}-\smfrac{3}{2}S_{ad}\delta q^d ~,
\end{equation}
rather than the similar equation with $\delta q^d = 0 = {\cal C}^{2b}$
used by Maartens\cite{Maa97}.

Applying the same method to $\delta H_{ab}$,
$\delta {\rm Bian}(\dot{H}_{ab})$ would give
\begin{equation}
 0 = h_a{}^c h_b{}^d\dot{\delta H}_{cd} + \Theta \delta H_{ab} -
 3\sigma_{c<a}\delta H_{b>}{}^c - \half
 \eta_{cd(a}\sigma^c{}_{b)}\delta q^d~,
\end{equation}
(in which the possible $\delta E_{bc}$ term is taken as zero due to
$\delta \pi_{ab} = 0$ and $\delta(E_{ab}-\half\pi_{ab})=0$). Thus
\begin{equation}
 0 = \dot{\cal C}^2_{ab} + \Theta {\cal C}^2_{ab} -
 3\sigma_{c<a}{\cal C}^2_{b>}{}^c + \half
 \eta_{cd(a}\sigma^c{}_{b)}{\cal C}^{1d}~.
\end{equation}
Maartens\cite{Maa97} has
\begin{equation}
 0 = \dot{\cal C}^2_{ab} + \Theta {\cal C}^2_{ab}  +
 \eta_{cd(a}\sigma^c{}_{b)}{\cal C}^{1d}~.
\end{equation}
This discrepancy is again due to an inconsistent way of taking a
commutator, which should in this context read, in Maartens' notation
in which $\curl S_{ab} = \eta_{cd(a}D^cS_{b)}{}^d$,
\begin{eqnarray}
(\curl S_{ab})\dot{} &=& \curl \dot{S}_{ab} - \smfrac{1}{3}\Theta
\curl S_{ab} - \sigma_{e}{}^{c}\eta_{cd(a}D^eS_{b)}{}^d \nonumber \\
&& +3(H_{c<a}+{\cal C}^2_{c<a})S_{b>}{}^c
+\half \eta_{cd(a}S^c{}_{b)}{\cal C}^{1d}~, 
\end{eqnarray}
Velden has
\begin{equation}
 0 = \dot{\cal C}^2_{ab} + \Theta {\cal C}^2_{ab}  +
 \smfrac{3}{2}\eta_{cd(a}\sigma^c{}_{b)}{\cal
 C}^{1d}~,
\end{equation}
but the source of this discrepancy is not known to me.

It is intended to give the remaining calculations for the irrotational
dust case, and the results for more general
cases, elsewhere.

\section*{Acknowledgements}
I am grateful to Profs.\ Brian Edgar and George Ellis, and to Drs.\ Roy
Maartens, Alan Rendall and Henk van Elst, for helpful discussions, and
to Edgar, Ellis, Maartens, van Elst and Prof.\ J\"urgen Ehlers for
copies of relevant references.


\section*{References}
\begin{thebibliography}{10}

\bibitem{Els96}
H.~{van Elst},
Ph.D. thesis, Queen Mary and Westfield College,
  London (1996).

\bibitem{Maa97}
R.~Maartens,
{\em Phys. Rev. D} {\bf 55}, 463 (1997).

\bibitem{MaaLesEll97}
R.~Maartens, W.M. Lesame and G.F.R. Ellis,
{\em Class. Quant. Grav.} {\bf 15}, 1005 (1998).

\bibitem{Vel97}
T.~Velden,
Diplomarbeit, University of Bielefeld (1997).

\bibitem{Edg80}
S.B. Edgar, {\em Gen. Rel. Grav.} {\bf 12}, 347 (1980).

\bibitem{Pap70}
A.~Papapetrou, {\em Ann. Inst. H. Poincar\'e} {\bf 13}, 271 (1970).

\bibitem{Pap71}
A.~Papapetrou, {\em C. R. Acad. Sci. (Paris) A} {\bf 272}, 1537 (1971).

\bibitem{Pap71a}
A.~Papapetrou, {\em C. R. Acad. Sci. (Paris) A}, {\bf 272}, 1613 (1971).

\bibitem{KraSteMac80}
D.~Kramer, H.~Stephani, M.A.H. MacCallum and E.~Herlt,
{\em Exact solutions of {Einstein's} field equations}
(Deutscher Verlag der Wissenschaften, Berlin, and Cambridge University
  Press, Cambridge, 1980).

\bibitem{Cha83}
S.~Chandrasekhar,
{\em The mathematical theory of black holes}.
(Oxford University Press, Oxford, 1983).

\bibitem{Edg92}
S.B. Edgar, {\em Gen. Rel. Grav.} {\bf 24}, 1267 (1992).

\bibitem{Ehl61}
J.~Ehlers, {\em Akad. Wiss. Lit. Mainz, Abh. Math.-Nat. Kl.} 11 (1961).
(English translation by G.F.R. Ellis and P.K.S. Dunsby, {\em Gen.
  {R}el. {G}rav.} {\bf 25}, 1225 (1993)).

\bibitem{Ell71}
G.F.R. Ellis in {\em General relativity and cosmology}, ed.\ R.K. Sachs,
  volume~47 of {\em Proceedings of the International School of Physics `Enrico
  Fermi'} (Academic Press, New York and London, 1971).

\bibitem{ElsUggLes97}
H.~{van Elst}, C.~Uggla, W.M. Lesame, G.F.R. Ellis and R.~Maartens,
{\em Class. Quant. Grav.} {\bf 14}, 1151 (1997).

\bibitem{Sop97}
C.F. Sopuerta, {\em Phys. Rev. D} {\bf 55}, 5936 (1997).

\bibitem{JanCarBin92}
R.T. Jantzen, P.~Carini and D.~Bini, {\em Ann. Phys. (N.Y.)} {\bf
215}, 1 (1992).

\bibitem{MacSke94}
M.A.H. MacCallum and J.E.F. Skea
in {\em Algebraic
  computing in general relativity (Proceedings of the first Brazilian school on
  computer algebra, vol 2)}, ed.\ M.J. Rebou\c{c}as and W.L. Roque (Oxford
  University Press, Oxford, 1994).

\end{thebibliography}
\end{document}